\documentclass[letterpaper, 10 pt, conference]{ieeeconf}  %

\IEEEoverridecommandlockouts                              %
\usepackage{tikzsymbols} %
\usepackage{float}    
\usepackage{graphicx}    
\usepackage{color}
\usepackage{epstopdf} 
\usepackage{graphicx}
\usepackage{amsfonts,amsmath,amssymb}
\usepackage{multirow}
\usepackage{mathtools}
\usepackage[most]{tcolorbox}
\usepackage{xcolor}
\usepackage{adjustbox}
\usepackage{epstopdf}
\usepackage{amsmath}
\usepackage{float}
\usepackage{epsfig}
\usepackage{tabularx}
\usepackage{cite}
\usepackage{booktabs}
\usepackage{siunitx}
\usepackage{placeins}
\definecolor{greenbold}{HTML}{015800}

\newcommand{\tss}[1]{\textsuperscript{#1}}
\newcommand*\crule[3]{%
  {\color[rgb]{#1}\rule[0.4ex]{#2}{#3}}}

\DeclarePairedDelimiterX{\norm}[1]{\lVert}{\rVert}{#1}
\title{\LARGE \bf
On the reduction of Linear Parameter-Varying State-Space models
}

\author{%
    E. Javier Olucha\tss{1}, Bogoljub Terzin\tss{1}, Amritam Das\tss{1}, Roland T\'oth\tss{1,2}\\ 
    {\normalsize\tss{1} Control Systems Group, Eindhoven University of Technology, The Netherlands.}\\
    {\normalsize\tss{2} Systems and Control Laboratory, Institute for Computer Science and Control, Budapest, Hungary.} \\
    {\normalsize\texttt{\{e.j.olucha.delgado, am.das, r.toth\}@tue.nl \& b.terzin@student.tue.nl}  \vspace{-10pt}}
}
\date{\today}
\let\svthefootnote\thefootnote
\newcommand{\thx}{\vspace{-10pt} \let\thefootnote\relax\footnotetext{This work has been supported by The MathWorks Inc. and by the European Union within the framework of the National Laboratory for Autonomous Systems (RRF-2.3.1-21-2022-00002). Opinions, findings, conclusions or recommendations expressed in this abstract are those of the authors and do not necessarily reflect the views of The MathWorks Inc. or the European Union.}
\let\thefootnote\svthefootnote}

\begin{document}

\maketitle\thx
\thispagestyle{empty}
\pagestyle{empty}

\begin{abstract}
This paper presents an overview and comparative study of the state of the art in \emph{state-order reduction} (SOR) and \emph{scheduling dimension reduction} (SDR) for \emph{linear parameter-varying} (LPV) \emph{state-space} (SS) models, comparing and benchmarking their capabilities, limitations and performance. The use case chosen for these studies is an interconnected network of nonlinear coupled mass spring damper systems with three different configurations, where some spring coefficients are described by arbitrary user-defined static nonlinear functions. For SOR, the following methods are compared: \emph{linear time-invariant} (LTI), LPV and LFR-based balanced reductions, moment matching (MM) and parameter-varying oblique projection. For SDR, the following methods are compared: \emph{principal component analysis} (PCA), trajectory PCA, Kernel PCA and LTI balanced truncation, autoencoders and deep neural network. The comparison reveals the most suitable reduction methods for the different benchmark configurations, from which we provide use case SOR and SDR guidelines that can be used to choose the best reduction method for a given LPV-SS model.
\end{abstract} \vspace{-8pt}

\section{Introduction} \vspace{-8pt}
Recently, the \emph{linear parameter-varying} (LPV) framework has been receiving a significant amount of attention for analysis and controller synthesis of \emph{nonlinear} (NL) and \emph{time-varying} (TV) systems. Beyond the realm of theoretical studies, this framework is widely applicable in diverse industrial context, as shown in \cite{hoffmannSurveyLinearParameterVarying2015}. Although most methods for analysis and controller synthesis of LPV models have been developed, one of the major bottlenecks of the framework is the transformation of a NL or TV model into an LPV one. This operation, commonly referred to as \emph{embedding}, is generally executed executed either by hand or by means of heuristic methods, which depends on the expertise of the user and often can lead to undesirably complex LPV models. While some procedures to automatize the LPV embedding have been recently developed \cite{kwiatkowskiAutomatedGenerationAssessment2006, tothModelingIdentificationLinear2010, decaignyInterpolationBasedModelingMIMO2011a, sadeghzadehAffineLinearParameter2020, sadeghzadehAutoencoderNeuralNetworks2023}, they can still yield LPV representations with a considerable model order. The complexity of LPV models becomes a problem when used for analysis and synthesis, as the available powerful computational tools rapidly scale in computational and memory load with the scheduling dimension and the model order, easily reaching hardware limitations. Hence, from a practical viewpoint, it is crucial to reduce the complexity of LPV models, where the complexity depends on the number of state variables, the number of scheduling variables, and the dependency of the coefficients in the representation of the scheduling variables. Moreover, the embedding operation typically introduces conservatism caused by the interdependence of the variations expressed by the scheduling variables. This aggregated conservatism affects the achievable performance of the synthesized controllers, therefore another objective of the LPV model reduction is to reduce this conservatism.

To address the complexity reduction problem in the LPV case, several reduction methods have been introduced. On the reduction of the number of state variables, \cite{moorePrincipalComponentAnalysis1981} introduces a balanced reduction for \emph{linear time-invariant} (LTI) models. A proper formulation under the scheduling variation is developed in \cite{woodApproximationLinearParametervarying1996}, which relies on finding static controllability and observability gramians for the LPV model. In \cite{beckCoprimeFactorsReduction2006}, a balanced reduction for \emph{linear fractional representation} (LFR) models is presented. From an alternative perspective, \cite{bastugMomentMatchingBased2015} proposes a moment matching method which approximates the input-output behaviour of affine LPV models. At last, \cite{theisLPVModelOrder2018} develops an extension of the Petrov-Galerkin approximation with a {parameter-varying} oblique projection.

On the reduction of the number of scheduling variables, a \emph{principal component analysis} (PCA) approach based on data from the scheduling trajectories is proposed in \cite{kwiatkowskiPCABasedParameterSet2008}. This method is enhanced in \cite{sadeghzadehAffineLinearParameter2020}, proposing a reduction based on the variation of the state space matrices. Another PCA approach is detailed in \cite{rizviKernelBasedPCAApproach2016}, where a kernel enables a nonlinear reduction mapping, followed by an optimization step to obtain the reduced LPV state-space matrices. Furthermore, the balanced reduction from \cite{moorePrincipalComponentAnalysis1981} can be used to reduce the scheduling dimension by interchanging the roles of the state and latent variables. Finally, the capabilities of \emph{neural networks} (NN) are explored in \cite{rizviModelReductionLinear2018}, proposing an \emph{autoencoder} (AE) NN to obtain a nonlinear reduction map, followed by an optimization step similar to \cite{rizviKernelBasedPCAApproach2016}. Based on this concept, \cite{koelewijnSchedulingDimensionReduction2020} proposes the use of a \emph{deep neural network} (DNN), able to directly determine the reduced LPV state-space matrices together with the scheduling reduction.

While there is an extensive number of procedures to tackle the LPV reduction problem, there is no overview present about the available methods. Furthermore, there are no comparative studies or reliable analysis of the strengths and weaknesses of the available methods, and it is difficult for the user to make the appropriate choices in practice. Therefore, as a contribution of this paper, we provide an overview and a comparison among the existing methods over different benchmarks that cover a broad range of use cases, and we determine the most suitable procedures based on different performance metrics.

The paper is structured as follows, in Section~\ref{sec: problem definition}, a mathematical problem definition of the LPV reduction problem is given. An overview of the existing LPV model reduction methods is given in Section~\ref{sec: overview}. Section~\ref{sec: benchmarks} describes the different benchmarks designed to compare the reduction methods. In Section~\ref{sec: results}, the performance comparison of the reduction methods is given. Finally, Section~\ref{sec: conclusion} draws the conclusions on the given results.

\emph{Notation:} The identity matrix of size $N$ is denoted by $I_N$. The matrix $A$ with matrices $A_1$ to $A_n$ on the diagonal is denoted by $A = \operatorname{diag}(A_1,\dots, A_n)$. The column vectorized form of a matrix $A \in \mathbb{R}^{m \times n}$ is denoted by $\Vec{A} \in \mathbb{R}^{m n}$, and the column vector $\left[x_1^\top~ \cdots ~ x_n^\top \right]^\top$ is denoted by $\operatorname{col}(x_1, \dots, x_n)$. The set of symmetric real matrices is denoted by $\mathbb{S}^n = \{ X \in \mathbb{R}^{n \times n} \; | \; X = X^\top\}$. A matrix $A \in \mathbb{S}^n$ is called positive-definite if $x^\top A x > 0, \; \forall  x \in \mathbb{R}^n \backslash \{ 0 \}$.
\vspace{-7pt}
\section{Problem definition \label{sec: problem definition}} \vspace{-8pt}
\subsection{LPV state-space embedding} \vspace{-5pt}
Consider a nonlinear system defined by the state-space representation:
\vspace{-10pt}
\begin{align}
\label{eq:NL_system}
\begin{split}
    \dot{x}(t) &= f(x(t),u(t)),  \\
    y(t) &= h(x(t),u(t)),
\end{split}
\end{align}
where $x(t): \mathbb{R} \rightarrow \mathbb{X} \subseteq \mathbb{R}^{n_x}$ is a state associated with (\ref{eq:NL_system}), while $u(t): \mathbb{R} \rightarrow \mathbb{U} \subseteq \mathbb{R}^{n_u}$ is the input, and  $y(t): \mathbb{R} \rightarrow \mathbb{Y} \subseteq \mathbb{R}^{n_y}$ is the output of the system. The functions $f: \mathbb{X} \times \mathbb{U} \rightarrow \mathbb{R}^{n_x}$ and $h: \mathbb{X} \times \mathbb{U} \rightarrow \mathbb{R}^{n_y}$ are assumed to be Lipschitz continuous. Embedding of the NL system (\ref{eq:NL_system}) in an LPV representation corresponds to constructing \cite{Toth2008thesis}:
\vspace{-5pt}
\begin{subequations}
\begin{align}
\label{eq:LPVSS}
    &\dot{x}(t) = A(p(t))x(t)+B(p(t))u(t), \nonumber\\
    &y(t) = C(p(t))x(t) + D(p(t))u(t), \\
    &\eta(x(t),u(t)) = p(t), \label{eq: eta_map}
\end{align}
\end{subequations}
where $p(t) \in \mathbb{P} \subseteq \mathbb{R}^{n_p}$ is the \emph{scheduling variable}, and $\eta$ is the \emph{scheduling map} that describes the internal dependence of scheduling variables on state and input values. 

However, different problems arise regarding the complexity of the model. First, the obtained LPV model might be of too high order for the utilization objective (e.g. control design). Additionally, the LPV embedding introduces conservatism by intentionally hiding the dependency of the $p(t)$ on $x(t)$ and $u(t)$, which should be reduced to get an effective representation of (\ref{eq:NL_system}). At last, regarding the complexity of the scheduling dependency, the embedding of (\ref{eq:LPVSS}) can easily result in polynomial or rational dependency on $p$, which is undesirable in control synthesis. Hence it is desirable to reduce the complexity of the dependence preferably to affine, while preserving the representation capability of (\ref{eq:LPVSS}).
\vspace{-10pt}
\subsection{State order reduction problem} \vspace{-5pt}
The \emph{state order reduction} (SOR) problem for LPV models consists of finding a \emph{reduced-order} (RO) LPV model approximation of (\ref{eq:LPVSS}) as:
\begin{subequations}
\begin{align}
\label{eq:LPVSS_states}
    &\dot{\hat{x}}(t) = \hat{A}(p(t))\hat{x}(t)+\hat{B}(p(t))u(t) \nonumber\\
    &\hat{y}(t) = \hat{C}(p(t))\hat{x}(t) + \hat{D}(p(t))u(t), \\
    &\eta(\hat{x}(t),u(t)) = \hat{p}(t), \label{eq: LPVSS_SOR_map}
\end{align}
\end{subequations}
where the reduced state vector $\hat{x}(t) \in \mathbb{R}^{r_x} \ll \mathbb{R}^{n_x}$.
\vspace{-10pt}
\subsection{Scheduling dimension reduction problem} \vspace{-5pt}
After the embedding procedure, the scheduling map $\eta$ can be high dimensional containing many nonlinear functions that are dependent on the same elements of $x$ and $u$. Therefore, there may exist a large dependency between the scheduling variables, which contributes to the conservatism of the LPV model. This can dramatically affect the capability of LPV synthesis to find a stabilizing controller with acceptable performance \cite{koelewijnSchedulingDimensionReduction2020}. The \emph{scheduling dimension reduction} (SDR) problem can be defined as follows:
given an LPV embedding (\ref{eq:LPVSS}) of the NL system (\ref{eq:NL_system}) and a set of nominal scheduling-variable trajectories of (\ref{eq:LPVSS}), denoted by $\mathcal{D}$ that corresponds to the typical behavior of (\ref{eq:NL_system}), find an LPV embedding in a form:
\vspace{-5pt}
\begin{subequations}
\begin{align}
\label{eq:LPVSS_scheduling}
    &\dot{\bar{x}}(t) = \bar{A}(\phi(t))\bar{x}(t)+\bar{B}(\phi(t))u(t), \nonumber\\
    &\bar{y}(t) = \bar{C}(\phi(t))\bar{x}(t) + \bar{D}(\phi(t))u(t), \\
    &\mu(p(t)) = \phi(t), \label{eq: LPVSS_redmap}
\end{align}
\end{subequations}
where $\phi(t) \in \Phi \subseteq \mathbb{R}^{n_{\phi}}$ is the reduced scheduling variable with $n_{\phi} \ll n_p$ such that (\ref{eq:LPVSS_scheduling}) approximates (\ref{eq:LPVSS}) for all trajectories $p \in \mathcal{D}$, and the set $\mathcal{D}$ introduced for (\ref{eq:LPVSS_scheduling}) is less conservative than (\ref{eq:LPVSS}). Here, approximation is considered in the sense that $\phi = \mu  \circ  \eta$, with $\mu: \mathbb{R}^{n_{p}} \rightarrow \mathbb{R}^{n_{\phi}}$, where
\begin{equation} \label{eq: Mbar}
    \bar{M}(\phi) = \begin{bmatrix}
        \bar{A}(\phi) & \bar{B}(\phi) \\ \bar{C}(\phi) & \bar{D}(\phi)
    \end{bmatrix} \approx M(p).
\end{equation}
If possible, also find the inverse mapping
\vspace{-7pt}
\begin{equation} \label{eq: inverse_map}
    \mu^{-1}(\phi(t)) = \bar{p}(t), \quad \text{s.t.} \quad \bar{p}(t) \approx p(t).   
\end{equation}
\subsection{Performance metrics \label{sub: performanceMetrics}} \vspace{-6pt}
We consider several performance metrics to assess the capabilities of the reduction methods from different viewpoints. First, we consider the \emph{computation time} (CPU) required to execute a reduction algorithm.

Next, let $\mathrm{NRMSE_{(\bullet)}}$ denote the \emph{Normalized Root-Mean-Square Error} of the simulated RO-LPV model w.r.t. the FO-LPV model, given as
\vspace{-7pt}
\begin{subequations}
\begin{align} \label{eq: NRMSE_SOR}
    &{\mathrm{NRMSE_{SOR}}} = \frac{\norm{{y} - \hat{y}}}{\norm{y - y_{\mathrm{mean}}}} \cdot 100, \\
    &{\mathrm{NRMSE_{SDR}}} = \frac{\norm{{y} - \bar{y}}}{\norm{y - y_{\mathrm{mean}}}} \cdot 100, \label{eq: NRMSE_SDR}
\end{align}
\end{subequations}
where $\{y(kT_\mathrm{s})\}_{k=0}^N$, $\{\hat{y}(kT_\mathrm{s})\}_{k=0}^N$ and $\{\bar{y}(kT_\mathrm{s})\}_{k=0}^N$ are the respective output signals from \eqref{eq:LPVSS}, \eqref{eq:LPVSS_states}, and \eqref{eq:LPVSS_scheduling} sampled at time instances $t = k T_\mathrm{s}$, and $y_{\text{mean}}$ is the mean of of the sampled sequence $y$.

Now, let $\lambda_{{2},(\bullet)}$ and $\lambda_{\infty,(\bullet)}$ denote the local $\mathcal{H}_2$ and $\mathcal{H}_\infty$ norm of the approximation error, given as
\vspace{-5pt}
\begin{align} \label{eq: lambda_SOR}
    &\lambda_{{2},\mathrm{SOR}}^{(i)} =  \norm{G_{p_\ast^{(i)}} - \hat{G}_{p_\ast^{(i)}}}_{2}, ~ \lambda_{\infty,\mathrm{SOR}}^{(i)} =  \norm{G_{p_\ast^{(i)}} - \hat{G}_{p_\ast^{(i)}}}_{\infty}, \\
    &\lambda_{{2},\mathrm{SDR}}^{(i)} =  \norm{G_{p_\ast^{(i)}} - \bar{G}_{p_\ast^{(i)}}}_{2}, ~ \lambda_{\infty,\mathrm{SDR}}^{(i)} =  \norm{G_{p_\ast^{(i)}} - \bar{G}_{p_\ast^{(i)}}}_{\infty}, \label{eq: lambda_SDR}
\end{align}
 where $G_{p_\ast^{(i)}}$, $\hat{G}_{p_\ast^{(i)}}$ and $\bar{G}_{p_\ast^{(i)}}$ are the respective local frequency response functions of \eqref{eq:LPVSS}, \eqref{eq:LPVSS_states} and \eqref{eq:LPVSS_scheduling} for a frozen (constant) $p_*^{(i)}$. We denote $\mathcal{P}$ (different from $\mathbb{P}$) as a grid that contains $n_g$ scheduling values $\{p_*^{(i)}\}_{i=1}^{n_g} \in \mathbb{P}$. 
\vspace{-5pt}
\section{Overview of the reduction techniques \label{sec: overview}} \vspace{-5pt}
\subsection{Preliminaries} \vspace{-5pt}
Consider the LPV embedding (\ref{eq:LPVSS}) of the NL system (\ref{eq:NL_system}) with the scheduling map \eqref{eq: eta_map} is given. An LPV-LFR form is obtained by introducing the latent variables $w_{\Delta} \in \mathbb{R}^{n_w}$ and $z_{\Delta} \in \mathbb{R}^{n_z}$, and disconnecting the LTI part of the model from the scheduling variables:
\begin{subequations}
\begin{equation}
\begin{bmatrix}
    \dot{x}(t) \\ z_{\Delta}(t) \\ y(t)
\end{bmatrix} = 
\begin{bmatrix}
    A & B_d & B_u \\
    C_d & D_{dd} & D_{du}\\
    C_y & D_{yd} & D_{yu}
\end{bmatrix}
\begin{bmatrix}
    x(t) \\ w_{\Delta}(t) \\ u(t)
\end{bmatrix},
    \label{lfr_G}
\end{equation}
together with the feedback path represented by:
\begin{equation}
    w_{\Delta}(t) = \Delta(p(t))z_{\Delta}(t),
    \label{eq:lpvlfr_delta}
\end{equation}
\end{subequations}
where $\Delta: \mathbb{P} \rightarrow \mathbb{R}^{n_w \times n_z}$ is the matrix function containing repeated instances of the scheduling variables. 
\vspace{-5pt}
\subsection{State-order reduction techniques} \vspace{-5pt}
\subsubsection{LTI balanced reduction (LTIBR) \label{subsub: LTI balred}}
One of the earliest SOR methods developed in \cite{moorePrincipalComponentAnalysis1981} for stable LTI models, is developed on the basis of considering only those states that are easy to reach (low control cost) and easy to observe (high observation energy), quantified by the controllability and observability gramians. For LPV models, it is applied by reducing \eqref{lfr_G}, in which the latent channels $z_{\Delta}$ and $w_{\Delta}$ are seen as additional output and input signals respectively. The underlying LTI model is transformed to an equivalent SS model using a balancing transformation, and thenceforth truncated to an approximated model that contains only the chosen number of the states. 
\subsubsection{LPV balred (LPVBR) \label{subsub: LPV balred}}
The first proper formulation of balanced reduction for LPV models where the scheduling dependence was taken into account has been developed in \cite{woodApproximationLinearParametervarying1996} for quadratically stable LPV models. This method relies on obtaining static (independent of $p$) controllability and observability gramians $P$ and $Q$, by solving the following \emph{linear matrix inequalities} (LMIs) constrained optimization problem on a gridded $p \in \mathcal{P}$:
\vspace{-5pt}
\begin{align}
\label{eq:LPV_gramians}
\min_{P,Q} \quad &\operatorname{trace}(PQ)\\ \vspace{-3pt}
\mathrm{s.t.} \quad & A(p)P + PA^{\top}(p) + B(p)B^{\top}(p) \prec 0, \nonumber \\
& A^{\top}(p)Q + QA(p) + C^{\top}(p)C(p) \prec 0, \nonumber\\
    &    \forall p \in \mathcal{P}. \nonumber
\end{align}
The gramians are used to find a balancing transformation which is applied to \eqref{eq:LPVSS}, and truncated to obtain \eqref{eq:LPVSS_states} with the chosen number of states.

\subsubsection{Moment matching (MM) \label{subsub: MM}}
The moment matching method is formulated for LPV-SS models with affine static dependence on the scheduling variable in \cite{bastugMomentMatchingBased2015}. This method preserves the first $N$ impulse response coefficients, and it is computed by finding the $N$-partial reachability and $N$-partial observability spaces, and transforming the matrix functions in \eqref{eq:LPVSS} accordingly. This method can also be used to obtain a minimal realization of LPV-SS models. 
\subsubsection{Parameter-Varying Oblique Projections (PVOP) \label{subsub: Oblique}}
In the oblique projections method \cite{theisLPVModelOrder2018}, the ideas of balanced reduction and generalized controllability and observability gramians are used and extended to find a reduced order model. Since static gramians restrict the search space, this alternative employs local gramian approximations on a gridded space of scheduling variables, which is later interpolated.

\subsubsection{LFR-based balred (LFRBR) \label{subsub: LFR balred}}
A model reduction method for LPV-LFR and uncertain models is developed in \cite{beckCoprimeFactorsReduction2006}, and relies on reducing \eqref{lfr_G} extended with the latent variables. The extended LTI model, defined as 
\begin{equation} \label{eq: Beck_LTI}
    \begin{bmatrix}
        y_1 \\ y
    \end{bmatrix} = \begin{bmatrix}
        \Tilde{A} & \Tilde{B}\\ \Tilde{C}&\Tilde{D}
    \end{bmatrix} \begin{bmatrix}
        u_1\\u
    \end{bmatrix},
    \end{equation}
where
$\Tilde{A} = \begin{bmatrix}
    A_0 & B_d \\ C_z & D_{dd}
\end{bmatrix}, \Tilde{B} = \begin{bmatrix}
    B_u \\ D_{du}
\end{bmatrix}, \Tilde{C} = \begin{bmatrix}
    C_y & D_{yd}
\end{bmatrix}$, $\Tilde{D} = D_{yu}$, $y_1 = \operatorname{col}(\dot{x}, z_{\Delta})$, and $u_1 = \operatorname{col}(x, w_{\Delta})$, is required to be controllable and observable. Structured gramians for \eqref{eq: Beck_LTI} are obtained, used to find a balancing transformation, that can be used to truncate the dimensions of $\Tilde{A}$. The method can be extended to handle unstable models by using a right-coprime factorization of \eqref{eq:LPVSS}, see \cite{beckCoprimeFactorsReduction2006}.
\vspace{-5pt}
\subsection{Scheduling reduction techniques} \vspace{-5pt}
First, we define relevant notation related to the SDR methods. The nominal scheduling trajectories are obtained by simulation, sampled at time instances $t = kT_s, \, k = 0,\dots, N-1$, and collected in
\vspace{-3pt}
\begin{equation}
\label{data_matrix}
    \Gamma = \mathcal{N} \left(\begin{bmatrix}
        p(0) & \hdots & p((N-1)T_s)
    \end{bmatrix}\right) ~\in \mathbb{R}^{n_{p} \times N}
\end{equation}
and $\mathcal{D} = \{p(kT_s)\}_{k=0}^{N-1}$, where $\mathcal{N}$ is an affine centering and normalization function. Moreover, the matrices of \eqref{eq:LPVSS} can be joined as:
\vspace{-8pt}
\begin{equation}
\label{LPV_matrices}
    L(p) = \begin{bmatrix}
        A(p(t)) & B(p(t)) \\
        C(p(t)) & D(p(t))
    \end{bmatrix},
\end{equation}
and the data can be collected as the vectorized values of matrix $L(p)$ across the nominal trajectory
\begin{equation}
\label{data_matrix_traj}
    \Gamma_m =  \mathcal{N}\left(\begin{bmatrix}
        \overrightarrow{L}(p(0)) & \hdots & \overrightarrow{L}(p((N-1)T_s)
    \end{bmatrix}\right).
\end{equation}
\subsubsection{Principal Component Analysis (PCA) \label{subsub: PCA}}
The standard PCA approach, proposed in \cite{kwiatkowskiPCABasedParameterSet2008}, is one of the earliest SDR methods, and it applies \emph{singular value decomposition} (SVD) on $\Gamma$ \eqref{data_matrix} to find the principal components of the scheduling trajectories and projects them to a subspace of lower order. Using the \emph{economic SVD}:
\vspace{-5pt}
\begin{align}
    \Gamma = \begin{bmatrix}
        U_s & U_r
    \end{bmatrix}
    \begin{bmatrix}
        \Sigma_s & 0 \\ 
        0 & \Sigma_r 
    \end{bmatrix}
    \begin{bmatrix}
        V_s^\top \\ V_r^\top
    \end{bmatrix},
\end{align}
where $U_s \in \mathbb{R}^{n_{p} \times n_{\phi}}, V_s \in \mathbb{R}^{N \times n_{\phi}}$ are unitary and {$\Sigma_s \succ 0 \in \mathbb{R}^{n_{\phi} \times n_{\phi}}$}, containing the highest $n_{\phi}$ singular values of $\Gamma$. The reduced  scheduling variable \eqref{eq: LPVSS_redmap} is given by $\phi(t) = U_s^{\top}p(t)$, and the inverse mapping \eqref{eq: inverse_map} is obtained by $\bar{p}(t) = U_s\phi(t)$. The matrices of \eqref{eq:LPVSS_scheduling} are then constructed using the same projection matrix $U_{s}$.  

\subsubsection{Trajectory PCA (TPCA) \label{subsub: Trajectory PCA}}
This approach, developed in \cite{sadeghzadehAffineLinearParameter2020}, finds a reduced order LPV-SS model of an affine LPV-SS model with minimal overbounding. This method applies a PCA to the state-space matrices over the course of $\Gamma_m$ \eqref{data_matrix_traj}. The advantage over the standard PCA method is that it captures not only the scheduling variables, but also all the variations of the state-space matrices.
\subsubsection{Kernel PCA (KPCA) \label{subsub: KPCA}}
The KPCA method \cite{rizviKernelBasedPCAApproach2016} extends the PCA method by allowing a nonlinear mapping of the scheduling map. In the KPCA method, the data is mapped with a kernel function to a higher dimensional feature space, on which a PCA is applied. Choosing a kernel function $k: \mathbb{P} \times \mathbb{P} \rightarrow \mathbb{R}$ results in the gram matrix $K_{ij} = (\Theta(p(i))^\top \Theta(p(j))) = k(p(i), p(j)$,
where $K \in R^{N \times N}$ and $\Theta$ denotes the nonlinear mapping. However, this method requires an additional optimization step to obtain the state-space matrices in \eqref{eq:LPVSS_scheduling}.
\subsubsection{SDR balred (SDRBR) \label{subsub: SDR balred}}
The LTI balancing truncation method in \S~\ref{subsub: LTI balred} can be used for SDR by interchanging the roles of the states and scheduling variables of \eqref{lfr_G},-\eqref{eq:lpvlfr_delta}. The alternate representation 
connects the time-shift operator between the latent variables $w_{\Delta}$ and $z_{\Delta}$, resulting in the following LTI-SS model:
\vspace{-8pt}
\begin{subequations} \label{eq: SDR_balred}
\begin{align} 
    \dot{z} &= D_{dd} \hat{z} + \begin{bmatrix}
        C_d & D_{du}
    \end{bmatrix} \Tilde{u},\\
    \Tilde{y} &= \begin{bmatrix}
        B_d \\ D_{yd}
    \end{bmatrix} \hat{z} + \begin{bmatrix}
        A & B_u \\ C_y & D_{yu}
    \end{bmatrix}\Tilde{u},
\end{align}
\end{subequations}
with  $\Tilde{u} = \operatorname{col}(x,u)$, and $\Tilde{y} = \operatorname{col}(\dot{x},y)$. The balanced truncation is performed on \eqref{eq: SDR_balred} to reduce the dimension of the state $z$, therefore reducing the $\Delta$ block. In LPV-LFR forms, $\Delta$ contains repeated instances of the scheduling variables in the diagonal, hence its reduction means reducing the multiplicity of the scheduling variables and the scheduling dimension. 
\subsubsection{Autoencoders (AE) \label{subsub: AE}}
The AE method \cite{rizviModelReductionLinear2018} makes use of an AE NN, which is trained with \eqref{data_matrix} to construct the nonlinear mappings \eqref{eq: LPVSS_redmap} and \eqref{eq: inverse_map}. This is an advantage over KPCA in \S~\ref{subsub: KPCA} since both $\mu$ and $\mu^{-1}$ are co-synthesized. The AE NN consists of an encoder layer mapping $p$ to $\phi$ and a decoder layer mapping $\phi$ to $\bar{p}$. However, as in the KPCA method, a separate optimization procedure is required to obtain the reduced order matrices in \eqref{eq:LPVSS_scheduling}.
\subsubsection{Deep neural network (DNN) \label{subsub: DNN}}
The DNN approach \cite{koelewijnSchedulingDimensionReduction2020} provides a method that allows a better approximation of more complex scheduling maps. This method, like the AE method, uses an NN to construct the mapping $\phi = \mu(p)$, with the benefit that the output of the network is a vectorized form $\hat{M}(p)$ \eqref{eq: Mbar} so that the reduced order matrices are directly constructed, avoiding the optimization procedure.
\vspace{-10pt}
\section{Benchmark definition \label{sec: benchmarks}} \vspace{-4pt}
\subsection{Interconnected Mass-Spring-Damper systems} \vspace{-5pt}
\begin{table}[t] \vspace{-10pt}
    \centering
    \caption{Affine LPV-SS model properties of the MSD configurations} \vspace{-5pt} \label{tab: MSD_config} 
    \begin{tabularx}{0.9\columnwidth} { 
     >{\raggedleft\arraybackslash}X >{\raggedleft\arraybackslash}X 
     >{\raggedleft\arraybackslash}X >{\raggedleft\arraybackslash}X 
     >{\raggedleft\arraybackslash}X  }
     \toprule
     & $N_m$ & $n_x$ & $n_{p}$\\
    \midrule
    $\mathrm{MSD}_1$ & 5 & 10 & 9 \\
    $\mathrm{MSD}_2$ & 50 & 100 & 3 \\
    $\mathrm{MSD}_3$ & 50 & 100 & 99 \\
    \bottomrule
    \end{tabularx} \vspace{-5pt}
\end{table} 
The benchmark model we consider is an interconnection of \emph{mass-spring-damper} (MSD) systems, inspired by \cite{mathworksReducedOrderModeling2023}, where a MSD connected to a rigid wall is governed by the following nonlinear continuous time state-space representation:
\begin{subequations}
\begin{align}
    &\frac{d}{dt} \begin{bmatrix}
        x(t) \\ \dot{x}(t)
    \end{bmatrix} = \begin{bmatrix}
        \dot{x}(t) \\
        \frac{1}{m} \left( \, u(t) - b(\dot{x}(t)) - k(x(t)) \, \right)
    \end{bmatrix}, \\
    \label{eq: MSD_coeff}
    &k(x(t)) = k_1 x(t) + k_2 x^3(t), \quad b(\dot{x}(t)) = b_1 \dot{x}(t),
\end{align}
\end{subequations}
where $x(t)$ and $\dot{x}(t)$ are the position and velocity of the mass in meters and meters per second, respectively, $m = 1 \mathrm{kg}$ is the mass, and $k_1 = 0.5 \mathrm{N/m}$, $k_2 = 1 \mathrm{N/m}$ and $b_1 = 1 \mathrm{Ns/m}$ are the respective spring and damper coefficient.

Now, we define 3 different MSD interconnection configurations denoted by $\mathrm{MSD_1}$, $\mathrm{MSD_2}$, $\mathrm{MSD_3}$, where a spring and damper with the coefficients~\eqref{eq: MSD_coeff} connects the adjacent masses, the position of the last mass is the output $y(t)$ in meters of the system and an input force $u(t)$ in Newtons is only applied to this last mass. The interconnected MSD blocks have the same physical parameters, but a different number of MSD blocks $N_m$ is defined for each configuration. In addition, we define $\mathrm{MSD_2}$ with $k_2 = 0$ except for the last three MSD blocks. The properties of the resulting full order nonlinear state-space models are displayed in Table~\ref{tab: MSD_config}. Then, these models are converted into affine LPV-SS models using the LPVcore toolbox (v.0.10), see \cite{boefLPVcoreMATLABToolbox2021}, functions \texttt{nlss} and \texttt{nlss2lpv}. By the implemented conversion approach, the number of scheduling variables $n_{p}$ is equal to the number of nonlinear springs.
\begin{figure}[t]
    \centering
    \begin{minipage}{0.48\columnwidth}
        \centering
        \includegraphics[width=0.95\columnwidth]{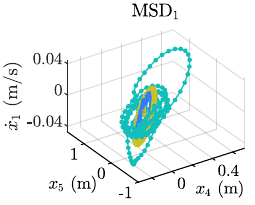} \vspace{-5pt}
            \caption{Visualization of the training and validation data with $u_{\mathrm{train}}(t)$ (\crule{0.2021, 0.4788, 0.9911}{3pt}{2pt}), $u_{\mathrm{in}}(t)$ (\crule{0.7858, 0.7573, 0.1598}{3pt}{2pt}) and $u_{\mathrm{out}}(t)$ 
            (\crule{0.0704, 0.7457, 0.7258}{3pt}{2pt}).}
            \label{fig:input_design}
    \end{minipage}\hfill
    \begin{minipage}{0.48\columnwidth}
        \centering
        \includegraphics[width=0.95\columnwidth]{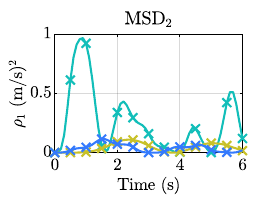} \vspace{-5pt}
        \caption{Visualization of the training grid $\mathcal{P}_{\mathrm{train}}$ ($\mathcolor[rgb]{0.2021, 0.4788, 0.9911}{\times}$) and the validation grids $\mathcal{P}_{\mathrm{in}}$($\mathcolor[rgb]{0.7858, 0.7573, 0.1598}{\times}$)  and $\mathcal{P}_{\mathrm{out}}$ ($\mathcolor[rgb]{0.0704, 0.7457, 0.7258}{\times}$).}\label{fig: grid_design}
    \end{minipage} \vspace{-14pt}
\end{figure}  
\vspace{-5pt}
\subsection{Input design for data-driven methods} \vspace{-5pt}
\begin{figure}[t]
    \centering
        \includegraphics[width=1\columnwidth]{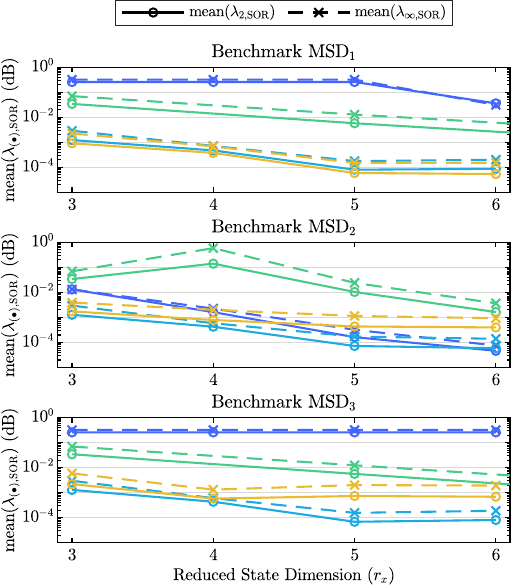} \vspace{-10pt} 
            \caption{Mean of the local $\mathcal{H}_2$ and $\mathcal{H}_\infty$ errors on $\mathcal{P}_{\mathrm{in}}$ with LTI balred (\crule{0.2647,0.4030, 0.9935}{6pt}{1pt}), LPV balred (\crule{0.1085,0.6669,0.8734}{6pt}{1pt}), Moment Matching (\crule{0.2809,0.7964,0.5266}{6pt}{1pt}) and Oblique Projections (\crule{0.9184,0.7308,0.1890}{6pt}{1pt}).}
            \label{fig: SOR_local_H_norms_IN}
\end{figure}
\begin{figure}[t!]
    \centering
        \includegraphics[width=1\columnwidth]{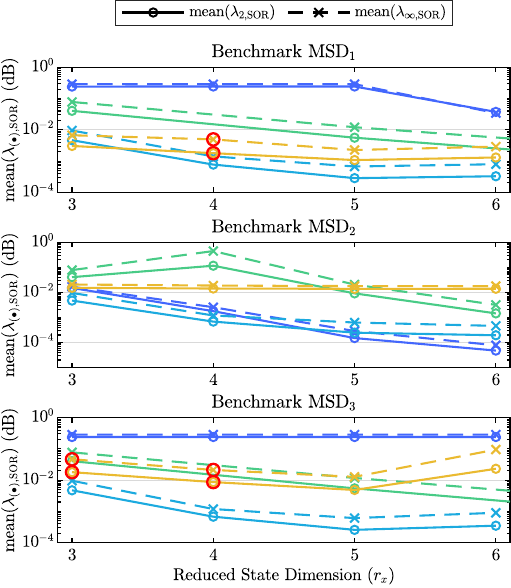} \vspace{-10pt} 
            \caption{Mean of the local $\mathcal{H}_2$ and $\mathcal{H}_\infty$ errors on $\mathcal{P}_{\mathrm{out}}$ with LTI balred (\crule{0.2647,0.4030, 0.9935}{6pt}{1pt}), LPV balred (\crule{0.1085,0.6669,0.8734}{6pt}{1pt}), Moment Matching (\crule{0.2809,0.7964,0.5266}{6pt}{1pt}) and Oblique Projections (\crule{0.9184,0.7308,0.1890}{6pt}{1pt}), where the unstable local model dynamics are excluded from the mean computation and indicated with \textcolor{red}{\large$\circ$}.}
            \label{fig: SOR_local_H_norms_OUT}
\end{figure}

The SDR methods require gathering system trajectories, but in practice, only a subset from the input and state-space, denoted by $\Pi_p \subset \mathbb{X} \times \mathbb{U}$, can be observed during experiments. Consequently, we propose distinct input sequences to train and validate the data-driven reduction methods, as well as to test their extrapolation capability. First, we define a training sequence $u_{\mathrm{train}}(t) \in \mathbb{U}_{\mathrm{train}}\subset \mathbb{U}$. Next, a first validation sequence is defined as $u_{\mathrm{in}}(t) \in \mathbb{U}_{\mathrm{in}} \subset \mathbb{U}, \mathbb{U}_{\mathrm{in}} \cap \mathbb{U}_{\mathrm{train}} = \emptyset$. Lastly, a second validation sequence is defined as $u_{\mathrm{out}}(t) \in \mathbb{U}_{\mathrm{out}} \subset \mathbb{U}$ subject to the restrictions $\mathbb{U}_{\mathrm{out}} \cap (\mathbb{U}\setminus \mathbb{U}_{\mathrm{in}}) \neq \emptyset$ and $ \mathbb{U}_{\mathrm{out}} \cap (\mathbb{U}\setminus \mathbb{U}_{\mathrm{train}}) \neq \emptyset$, as shown in Fig.~\ref{fig:input_design}.

Similarly, the methods described in \S~\ref{subsub: LPV balred} and \S~\ref{subsub: Oblique} rely on linearizations around a grid of scheduling points. First, we define a training grid $\mathcal{P}_{\mathrm{train}} \in \mathbb{P}_{\mathrm{train}}\subset \mathbb{P}$ of $n_g = 12$ scattered points. Next, a first validation grid  $\mathcal{P}_{\mathrm{in}} \in \mathbb{P}_{\mathrm{in}} \subset \mathbb{P}, \mathbb{P}_{\mathrm{in}} \cap \mathbb{P}_{\mathrm{train}} = \emptyset$ is defined. Lastly, we define a second validation grid $\mathcal{P}_{\mathrm{out}} \in \mathbb{P}_{\mathrm{out}} \subset \mathbb{P}$ subject to the restrictions $\mathbb{P}_{\mathrm{out}} \cap (\mathbb{P}\setminus \mathbb{P}_{\mathrm{in}}) \neq \emptyset$ and $ \mathbb{P}_{\mathrm{out}} \cap (\mathbb{P}\setminus \mathbb{P}_{\mathrm{train}}) \neq \emptyset$. Both $\mathcal{P}_{\mathrm{in}}$ and $\mathcal{P}_{\mathrm{out}}$ consist of $n_g = 21$ scattered points, and a visualization is shown in Fig.~\ref{fig: grid_design}.
\vspace{-7pt}
\section{Performance comparison results \label{sec: results}} \vspace{-6pt}
\subsection{State-order reduction \label{sub: SOR results}} \vspace{-5pt}
\sisetup{
tight-spacing=true,
text-series-to-math = true ,
propagate-math-font = true,
exponent-product =\ensuremath{\cdot},
print-zero-exponent = true
}
\begin{table}[t] \vspace{-10pt}
   \scriptsize
   \centering
    \caption{SOR performance for MSD1 with $r_x = 5$, $u_{\mathrm{out}}$ and $\mathcal{P}_{\mathrm{out}}$ } \vspace{-5pt} \label{tab: SOR_MSD1}
      \begin{tabularx}{1\columnwidth} { 
     >{\raggedright\arraybackslash}X >{\raggedright\arraybackslash}X 
     >{\raggedright\arraybackslash}X >{\raggedright\arraybackslash}X 
     >{\raggedright\arraybackslash}X  } 
     \toprule
    Metric & LTIBR & LPVBR & MM & PVOP \\
    \midrule
CPU (s) & \textbf{\textcolor{greenbold}{\num{4.28e-03}}} & \num{1.13} & \num{2.96e-01} & \num{5.10e-01}   \\
$\mathrm{NRMSE_{SOR}}$ & \num{7.97e+01} & \num{3.29e+01} & \textbf{\textcolor{greenbold}{\num{5.23e0}}} & $\infty$          \\
$\max(\lambda_{2,\mathrm{SOR}})$ & \num{2.69e-01} & \textbf{\textcolor{greenbold}{\num{6.98e-04}}} & \num{1.45e-03} & \num{3.39e-03}   \\
$\mathrm{std}(\lambda_{2,\mathrm{SOR}})$ & \num{2.22e-02} & \num{2.48e-04} & \textbf{\textcolor{greenbold}{\num{2.04e-04}}} & \num{9.49e-04}   \\
$\max(\lambda_{\infty,\mathrm{SOR}})$ & \num{3.47e-01} & \textbf{\textcolor{greenbold}{\num{2.08e-03}}} & \num{3.36e-03} & \num{5.62e-03}  \\
$\mathrm{std}(\lambda_{\infty,\mathrm{SOR}})$ & \num{4.44e-02} & \num{6.23e-04} & \textbf{\textcolor{greenbold}{\num{5.28e-04}}} & \num{1.74e-03} \\
    \bottomrule
    \end{tabularx}
    \caption{SOR performance for MSD2 with $r_x = 5$, $u_{\mathrm{out}}$ and $\mathcal{P}_{\mathrm{out}}$} \vspace{-5pt} \label{tab: SOR_MSD2} 
      \begin{tabularx}{1\columnwidth} { 
     >{\raggedright\arraybackslash}X >{\raggedright\arraybackslash}X 
     >{\raggedright\arraybackslash}X >{\raggedright\arraybackslash}X 
     >{\raggedright\arraybackslash}X  }  
     \toprule
    Metric & LTIBR & LPVBR & MM & PVOP \\
    \midrule
CPU (s) & \textbf{\textcolor{greenbold}{\num{1.69e-02}}}  & \num{1.45e+02}  & \num{ 2.12e-01  }  & \num{ 5.80e+01  }\\
$\mathrm{NRMSE_{SOR}}$ & \textbf{\textcolor{greenbold}{\num{ 2.26}}}  & $\infty$  & \num{ 4.14   }  & \num{2.45}\\
$\max(\lambda_{2,\mathrm{SOR}})$ & \textbf{\textcolor{greenbold}{\num{ 1.92e-04}}}  & \num{ 6.36e-04  }  & \num{ 1.22e-02  }  & \num{ 2.86e-01  }\\
$\mathrm{std}(\lambda_{2,\mathrm{SOR}})$ & \textbf{\textcolor{greenbold}{\num{3.23e-05}}}  & \num{ 2.18e-04  }  & \num{ 2.76e-03  }  & \num{ 6.24e-02  }\\
$\max(\lambda_{\infty,\mathrm{SOR}})$ & \textbf{\textcolor{greenbold}{\num{ 3.78e-04}}}  & \num{ 1.49e-03  }  & \num{ 2.83e-02  }  & \num{ 3.59e-01  }\\
$\mathrm{std}(\lambda_{\infty,\mathrm{SOR}})$ & \textbf{\textcolor{greenbold}{\num{ 7.74e-05}}}  & \num{ 5.62e-04  } & \num{ 7.29e-03} & \num{ 7.82e-02} \\
    \bottomrule
    \end{tabularx}
        \caption{SOR performance for MSD3 with $r_x=5$, $u_{\mathrm{out}}$ and $\mathcal{P}_{\mathrm{out}}$} \vspace{-5pt} \label{tab: SOR_MSD3}
      \begin{tabularx}{1\columnwidth} { 
     >{\raggedright\arraybackslash}X >{\raggedright\arraybackslash}X 
     >{\raggedright\arraybackslash}X >{\raggedright\arraybackslash}X 
     >{\raggedright\arraybackslash}X  }   
     \toprule
    Metric & LTIBR & LPVBR & MM & PVOP \\
    \midrule
CPU (s) & \textbf{\textcolor{greenbold}{\num{2.32e-02}}} & \num{ 1.04e+03 } & \num{ 4.58 } & \num{ 1.98e+01} \\
$\mathrm{NRMSE_{SOR}}$ & \num{8.26e+01 } & $\infty$ & \textbf{\textcolor{greenbold}{\num{ 3.26}}} & $\infty$        \\
$\max(\lambda_{2,\mathrm{SOR}})$ & \num{  2.69e-01 } & \textbf{\textcolor{greenbold}{\num{6.53e-04}}} & {\num{ 1.17e-03 }} & \num{ 1.47e-02} \\
$\mathrm{std}(\lambda_{2,\mathrm{SOR}})$& \num{       2.36e-02 } & \num{2.33e-04} & \textbf{\textcolor{greenbold}{\num{ 1.57e-04}}} & \num{ 4.71e-03 }\\
$\max(\lambda_{\infty,\mathrm{SOR}})$ & \num{    3.49e-01 } & \textbf{\textcolor{greenbold}{\num{1.92e-03}}} &  {\num{2.68e-03 }} &  \num{4.14e-02} \\
$\mathrm{std}(\lambda_{\infty,\mathrm{SOR}})$ &   \num{4.66e-02 } &  \num{5.85e-04} &  \textbf{\textcolor{greenbold}{\num{4.08e-04}}} &  \num{1.32e-02} \\
    \bottomrule
    \end{tabularx} \vspace{-10pt}
\end{table}
The SOR methods have been implemented in the LPVcore toolbox (v.0.10), and are applied to the benchmarks proposed in Subsection~\ref{sec: benchmarks}. The resulting reduced order models are compared based on performance metrics proposed in Subsection~\ref{sub: performanceMetrics}. We first compare the mean of $\lambda_{(\bullet),\mathrm{SOR}}$ \eqref{eq: lambda_SOR} on $\mathcal{P}_{\mathrm{in}}$, as shown in Fig.~\ref{fig: SOR_local_H_norms_IN}. While \emph{LPV balred} and \emph{Oblique Projections} give the best results, \emph{Moment Matching} exhibits superior accuracy performance with respect to the other methods as we increase the dimension of the reduced order model. Lastly, \emph{LTI balred} displays high accuracy for $\mathrm{MSD_2}$, and this is because $\mathrm{MSD_2}$ is dominated by linear dynamics.

\begin{figure}[t]
    \centering
       \includegraphics[width=1\columnwidth]{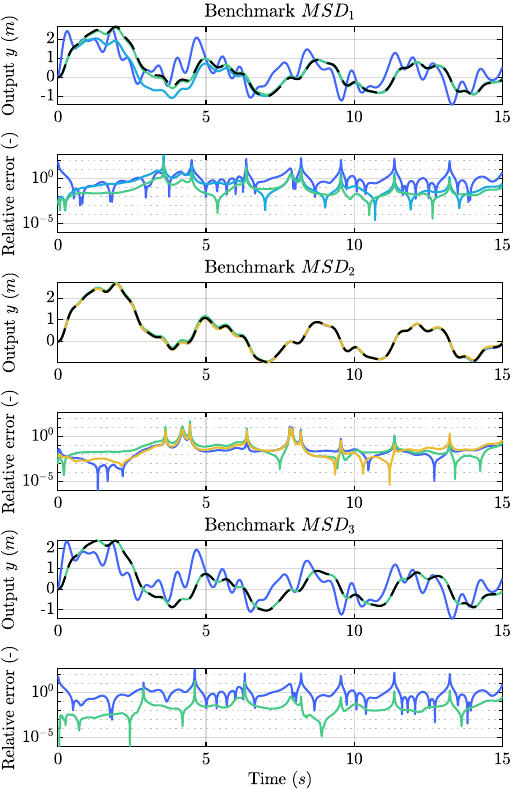} \vspace{-10pt} \caption{Self-scheduled simulation of the FO (\crule{0,0, 0}{3pt}{1pt} \crule{0,0, 0}{3pt}{1pt}) and RO models obtained with LTI balred (\crule{0.2647,0.4030, 0.9935}{6pt}{1pt}), LPV balred (\crule{0.1085,0.6669,0.8734}{6pt}{1pt}), Moment Matching (\crule{0.2809,0.7964,0.5266}{6pt}{1pt}) and Oblique Projections (\crule{0.9184,0.7308,0.1890}{6pt}{1pt}) for $r_x = 5$, $u_{\mathrm{out}}(t)$ and with zero initial conditions.}
            \label{fig: SOR_time_domain} \vspace{-20pt}
\end{figure}

Next, we test the extrapolation capabilities of the SOR methods by using $\mathcal{P}_{\mathrm{out}}$ to compare on the mean of $\lambda_{(\bullet),\mathrm{SOR}}$ \eqref{eq: lambda_SOR}, as shown in Fig.~\ref{fig: SOR_local_H_norms_OUT}. Due to the validation points outside the training space, there is an overall performance decrease compared to Fig.~\ref{fig: SOR_local_H_norms_IN}, and \emph{oblique projections} provided unstable local model dynamics at some points on $\mathcal{P}_{\mathrm{out}}$, which are indicated with a red circle in Fig.~\ref{fig: SOR_local_H_norms_OUT} and excluded from the computation of the mean. For this method, the worst-case ratio of unstable to stable local dynamics is of $9.5$ \% on the $\mathrm{MSD_1}$ and of $33.3$ \% for $\mathrm{MSD_3}$. Moreover, $u_{\mathrm{out}}(t)$ is used to run time-domain \emph{self-scheduled} simulations outside the training space, which shows the robustness of the methods. Only the stable simulations are shown in Fig.~\ref{fig: SOR_time_domain}, and \emph{Moment Matching} clearly displays the highest extrapolation capability. An overview of the performance obtained with $u_{\mathrm{out}}(t)$ and $\mathcal{P}_{\mathrm{out}}$ is provided in Tables~\ref{tab: SOR_MSD1}, \ref{tab: SOR_MSD2}, and \ref{tab: SOR_MSD3}. Note that, for the \emph{Oblique Projections}, the interpolation of the reduced models is included in the computation time, and the red numbers indicate that some local unstable models were excluded to compute the metrics.
\begin{table}[t] \vspace{-10pt}
    \small
    \centering
    \caption{Use case state order reduction guidelines} \vspace{-5pt} \label{tab: SOR_recommendations} 
    \begin{tabularx}{0.98\columnwidth} { 
     >{\hsize=.65\hsize }X >{\hsize=.35\hsize }X}
     \toprule
    Use case & Method\\
    \midrule
    Non-minimal or unstable model & Moment Matching \\
    Large $n_x$ with low $n_p$ & Oblique Projections \\
    Large $n_x$ with large $n_p$ & Moment Matching \\
    Higher accuracy at frozen points & LPV balred \\
    Dominant linear dynamics & LTI or LPV balred \\
    \bottomrule
    \end{tabularx} \vspace{-7pt}
\end{table}

Several limitations were found with some approaches during the execution of the SOR. A strict limitation for \emph{LFR-based balred} is in that the LTI block described in~\eqref{eq: Beck_LTI} is required to be controllable and observable. In practice, this condition is hardly satisfied, and we were not able to test this method in the proposed benchmarks. Second, all gramian-based methods require the FO model to be in a minimal and stable SS representation, and \emph{Moment Matching} is the only SOR method able to handle non-minimal or unstable SS representations. Third, the \emph{Oblique Projections} method requires the interpolation of local reduced models to recover the RO-LPV model, but this becomes computationally infeasible for a high-dimensional scheduling space. This limitation was found with $\mathrm{MSD_1}$ and $\mathrm{MSD_3}$, and we used a least-squares-based interpolation with affine basis to circumvent the problem. Lastly, the method \emph{LPV balred} relies on finding static gramians, which can be infeasible for many LPV systems. Based on these observations, a set of guidelines are summarized in Table~\ref{tab: SOR_recommendations} to assist the choice of SOR reduction method in various scenarios.
\vspace{-3pt}
\subsection{Scheduling dimension reduction \label{sub: SDR results}} 
\begin{figure}[t]
    \centering
        \includegraphics[width=1\columnwidth]{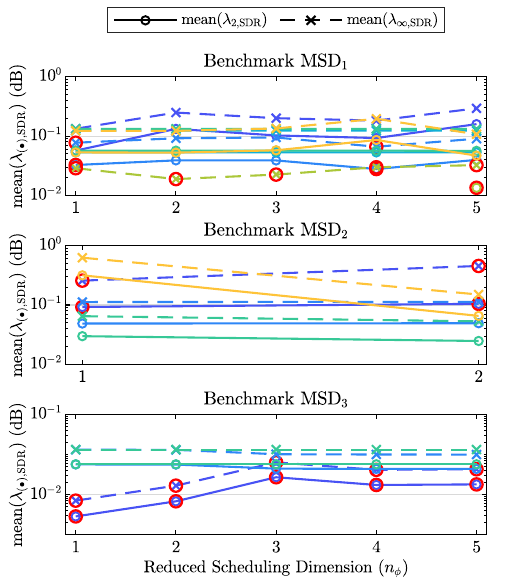}  \vspace{-5pt} 
            \caption{Mean of the local $\mathcal{H}_2$ and $\mathcal{H}_\infty$ errors on the validation grid $\mathcal{P}_{\mathrm{in}}$ of the methods PCA (\crule{ 0.2810,0.3228,0.9579}{6pt}{1pt}), Trajectory PCA (\crule{0.1786,0.5289,0.9682}{6pt}{1pt}), KPCA (\crule{0.0689,0.6948,0.8394}{6pt}{1pt}), \emph{SDR balred} (\crule{0.2161,0.7843,0.5923}{6pt}{1pt}), AE (\crule{0.6720,0.7793,0.2227}{6pt}{1pt}) and DNN (\crule{0.9970,0.7659,0.2199}{6pt}{1pt}), where the unstable local model dynamics are excluded from the mean computation and indicated with \textcolor{red}{\large$\circ$}.}
            \label{fig: SDR_local_H_norms_IN} \vspace{-15pt}
\end{figure}
\begin{figure}[t!]
    \centering
        \includegraphics[width=1\columnwidth]{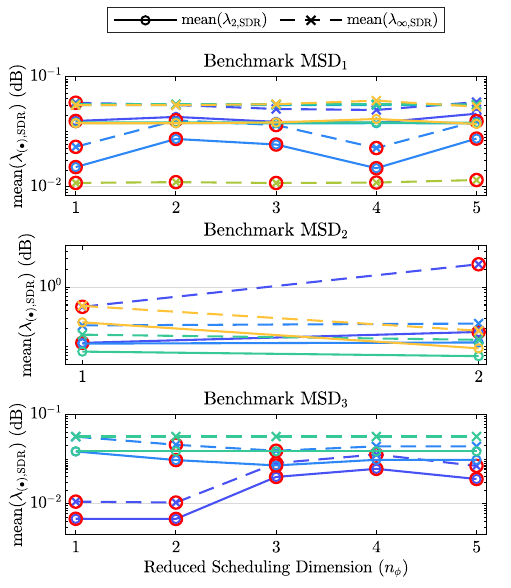} \vspace{-10pt} 
            \caption{Mean of the local $\mathcal{H}_2$ and $\mathcal{H}_\infty$ errors on the validation grid $\mathcal{P}_{\mathrm{out}}$ of the methods PCA (\crule{ 0.2810,0.3228,0.9579}{6pt}{1pt}), Trajectory PCA (\crule{0.1786,0.5289,0.9682}{6pt}{1pt}), KPCA (\crule{0.0689,0.6948,0.8394}{6pt}{1pt}), \emph{SDR balred} (\crule{0.2161,0.7843,0.5923}{6pt}{1pt}), AE (\crule{0.6720,0.7793,0.2227}{6pt}{1pt}) and DNN (\crule{0.9970,0.7659,0.2199}{6pt}{1pt}), where the unstable local model dynamics are excluded from the mean computation and indicated with \textcolor{red}{\large$\circ$}.}
            \label{fig: SDR_local_H_norms_OUT} \vspace{-15pt}
\end{figure}
\vspace{-8pt}
\begin{figure}[t]
    \centering
       \includegraphics[width=1\columnwidth]{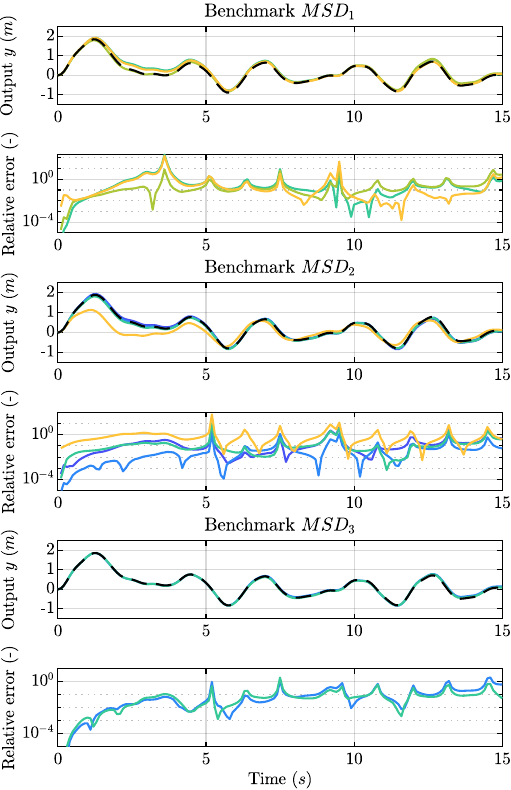}  \vspace{-10pt}  \caption{Self-scheduled simulation of the FO (\crule{0,0, 0}{3pt}{1pt} \crule{0,0, 0}{3pt}{1pt}) and RO models obtained with PCA (\crule{ 0.2810,0.3228,0.9579}{6pt}{1pt}), Trajectory PCA (\crule{0.1786,0.5289,0.9682}{6pt}{1pt}), \emph{SDR balred} (\crule{0.2161,0.7843,0.5923}{6pt}{1pt}), AE (\crule{0.6720,0.7793,0.2227}{6pt}{1pt}) and DNN (\crule{0.9970,0.7659,0.2199}{6pt}{1pt}) for $n_\phi = 1$, $u_{\mathrm{out}}(t)$ and with zero initial conditions.}
            \label{fig: SDR_time_domain} \vspace{-17pt}
\end{figure}
The SDR methods have been implemented in the LPVcore toolbox (v.0.10), and for \emph{KPCA}, the kernel function $k(p(i),p(j)) = \left((p(i),p(j)) + 1 \right)^d$ with $d=8$ is used. The \emph{DNN} method is applied on $\mathrm{MSD_1}$ with Levenberg-Marquardt backpropagation \cite{themathworksinc.DeepLearningToolbox2023}, and Scaled Conjugate Gradient is used on $\mathrm{MSD_2}$. Moreover, \emph{KPCA}, \emph{AE}, and \emph{DNN} were unable to reduce $\mathrm{MSD_3}$ because the laptop used had no sufficient memory, whereas the first two also failed to produce results for $\mathrm{MSD_2}$ after six hours of computation time.

Next, the mean of $\lambda_{(\bullet),\mathrm{SDR}}$ \eqref{eq: lambda_SDR} on $\mathcal{P}_{\mathrm{in}}$ is shown in Fig.~\ref{fig: SDR_local_H_norms_IN}. In this case, some of the local RO model dynamics are unstable at some points contained in $\mathcal{P}_{\mathrm{in}}$, which are indicated with a red circle and excluded from the computation of the mean. \emph{SDR balred} results in one of the most accurate methods, followed by \emph{Trajectory PCA} and \emph{DNN}. 

Now, the extrapolation capabilities of the SDR methods are tested. In Fig.~\ref{fig: SDR_local_H_norms_OUT}, the local error norm using $\mathcal{P}_{\mathrm{out}}$ is shown, were unstable local model dynamics are indicated with a red circle and excluded form the mean, and Table~\ref{tab: SDR_worstCase_ratio} displays the worst-case ratio of unstable to stable local error dynamics of the RO models. To follow, Fig.~\ref{fig: SDR_time_domain} shows a time-domain \emph{self-scheduled} simulation using $u_{\mathrm{out}}(t)$ and $\mathcal{P}_{\mathrm{out}}$. From this results, it is clear that \emph{SDR balred}, \emph{Trajectory PCA} and \emph{DNN} are able to predict the model outside the training data, and that their performance depends on the properties of the FO model. In addition, \emph{AE} was able to provide slightly better results than \emph{DNN} for $\mathrm{MSD_2}$, but \emph{DNN} is preferred as it is able to handle larger systems than \emph{AE}. Lastly, an overview of the performance obtained with $u_{\mathrm{out}}(t)$ and $\mathcal{P}_{\mathrm{out}}$ is given in Tables~\ref{tab: SDR_MSD1},~\ref{tab: SDR_MSD2}~and~\ref{tab: SDR_MSD3}.
 
The only limitations encountered are the computation resources required for \emph{AE}, \emph{KPCA} and \emph{DNN}, and that the \emph{SDR balred} requires~\eqref{eq: SDR_balred} to be a minimal and stable representation. Intuitively, the achievable performance of \emph{DNN} is significantly higher, provided more training data is available, or if applied to systems with more complex nonlinear behaviour. Based on these observations, a set of guidelines are summarized in Table~\ref{tab: SDR_recommendations} to assist the choice of SDR method in various scenarios.
\begin{table}[b] \vspace{-7pt}
    \small
    \centering
    \caption{Worst-case ratio in \% of unstable to stable local dynamics of the RO models obtained with SDR} \vspace{-5pt}\label{tab: SDR_worstCase_ratio}
    \begin{tabularx}{1\columnwidth} { 
     >{\hsize=.25\hsize }X >{\hsize=.25\hsize }X >{\hsize=.25\hsize }X >{\hsize=.25\hsize }X}
     \toprule
    Method & $\mathrm{MSD_1}$ & $\mathrm{MSD_2}$ & $\mathrm{MSD_3}$\\
    \midrule
    PCA & 23.8 & 57.1 & 81.0 \\
    TPCA & 61.9 & 0 & 33.3 \\
    AE & 85.7 & 0 & 0 \\
    \bottomrule
    \end{tabularx}
\end{table}
\begin{table*}[t] 
   \scriptsize 
   \centering
    \caption{SDR performance for MSD1 with $n_{\phi}=1$, $u_{\mathrm{out}}$ and $\mathcal{P}_{\mathrm{out}}$. Red indicates unstable local models were excluded from the metrics.} \vspace{-5pt} \label{tab: SDR_MSD1} 
      \begin{tabularx}{1.8\columnwidth} { 
     >{\raggedright\arraybackslash}X >{\raggedright\arraybackslash}X 
     >{\raggedright\arraybackslash}X >{\raggedright\arraybackslash}X 
     >{\raggedright\arraybackslash}X
     >{\raggedright\arraybackslash}X
     >{\raggedright\arraybackslash}X} 
     \toprule
    Metric & PCA & TPCA & KPCA & SDRBR & AE & DNN \\
    \midrule
CPU (s) & \textbf{\textcolor{greenbold}{\num{1.85e-01}}} & \num{3.77e-1} & \num{1.19e0} & \num{2.25e-1} & \num{1.69e0} & \num{5.77e2}  \\
$\mathrm{NRMSE_{SDR}}$ & $\infty$ & $\infty$ & $\infty$ & \num{1.89e1} & \textbf{\textcolor{greenbold}{\num{9.84e0}}} & {\num{1.51e1}}\\
$\max(\lambda_{2, \mathrm{SDR}})$ & \textcolor{red}{\num{4.89e-1}} & \textcolor{red}{\num{1.08e-1}} & \textbf{\textcolor{greenbold}{\num{4.75e-1}}} & \num{4.79e-1} & $\infty$ & \num{4.75e-1}\\
$\mathrm{std}(\lambda_{2,\mathrm{SDR}})$ & \textcolor{red}{\num{1.34e-1}} & \textcolor{red}{\num{3.68e-2}} & \num{1.28e-1} & \num{1.30e-1} & $\infty$ & \textbf{\textcolor{greenbold}{\num{1.28e-1}}}\\
$\max(\lambda_{\infty,\mathrm{SDR}})$ & \textcolor{red}{\num{9.31e-1}} & \textcolor{red}{\num{2.46e-1}} & \num{9.14e-1} & \num{9.28e-1} & \textcolor{red}{\num{3.04e-2}} & \textbf{\textcolor{greenbold}{\num{9.11e-1}}} \\
$\mathrm{std}(\lambda_{\infty,\mathrm{SDR}})$ & \textcolor{red}{\num{2.69e-1}} & \textcolor{red}{\num{8.39e-2}} & \num{2.56e-1} & \num{2.60e-1} & \textcolor{red}{\num{1.62e-2}} & \textbf{\textcolor{greenbold}{\num{2.55e-1}}}\\
    \bottomrule
    \end{tabularx} \vspace{-5pt}
\end{table*}
\begin{table}[t] 
    \scriptsize
    \centering
    \caption{SDR performance for MSD2 with $n_{\phi} = 1$, $u_{\mathrm{out}}$ and $\mathcal{P}_{\mathrm{out}}$} \vspace{-5pt}\label{tab: SDR_MSD2} 
      \begin{tabularx}{1\columnwidth} { 
     >{\raggedright\arraybackslash}X >{\raggedright\arraybackslash}X 
     >{\raggedright\arraybackslash}X >{\raggedright\arraybackslash}X 
     >{\raggedright\arraybackslash}X
     >{\raggedright\arraybackslash}X
     >{\raggedright\arraybackslash}X} 
     \toprule
    Metric & PCA & TPCA & SDRBR & DNN \\
    \midrule
CPU (s)                  & \num{1.24e-01} & \num{7.44e-1}       & \textbf{\textcolor{greenbold}{\num{9.80e-2}}}& \num{3.89e0} \\
$\mathrm{NRMSE_{SDR}}$& \num{7.39e0}   &\textbf{\textcolor{greenbold}{\num{2.75e0}}}& \num{6.80e0}          & \num{4.92e1} \\
$\max(\lambda_{2, \mathrm{SDR}})$        & \textcolor{red}{\num{4.11e-1}}            & \num{3.58e-1}       & \textbf{\textcolor{greenbold}{\num{2.98e-1}}}& \num{3.63e-1}\\
$\mathrm{std}(\lambda_{2,\mathrm{SDR}})$      & \textcolor{red}{\num{1.60e-1}}            & \num{1.02e-1}       & \num{8.49e-2}         & \textbf{\textcolor{greenbold}{\num{8.4e-2}}} \\
$\max(\lambda_{\infty,\mathrm{SDR}})$   & \textcolor{red}{\num{2.43e0}}            & \num{6.03e-1}       & \textbf{\textcolor{greenbold}{\num{5.39e-1}}}& \num{7.59e-1} \\
$\mathrm{std}(\lambda_{\infty,\mathrm{SDR}})$ & \textcolor{red}{\num{8.01e-1}}            & \num{1.88e-1}       & \textbf{\textcolor{greenbold}{\num{1.60e-1}}}& \num{2.03e-1} \\
    \bottomrule
    \end{tabularx}
          \caption{SDR performance for MSD3 with $n_{\phi} = 1 $, $u_{\mathrm{out}}$ and $\mathcal{P}_{\mathrm{out}}$} \vspace{-5pt} \label{tab: SDR_MSD3}
    \begin{tabularx}{1\columnwidth} { 
     >{\raggedright\arraybackslash}X >{\raggedright\arraybackslash}X 
     >{\raggedright\arraybackslash}X >{\raggedright\arraybackslash}X} 
     \toprule
    Metric & PCA & TPCA & \emph{SDRBR}\\
    \midrule
CPU (s)                 & \num{1.51e0}& \num{6.16e0}           & \textbf{\textcolor{greenbold}{\num{3.67e-1}}}\\
$\mathrm{NRMSE_{SDR}}$ & $\infty$         & \textbf{\textcolor{greenbold}{\num{6.08e0}}}  &\num{7.1e0}\\
$\max(\lambda_{2, \mathrm{SDR}})$       & \textcolor{red}{\num{8.82e-3}}         & \textbf{\textcolor{greenbold}{\num{4.77e-1}}} & \num{4.79e-1}\\
$\mathrm{std}(\lambda_{2,\mathrm{SDR}})$     & \textcolor{red}{\num{4.27e-3}}         &\textbf{\textcolor{greenbold}{\num{1.30e-1}}}  & \num{1.30e-1}\\
$\max(\lambda_{\infty,\mathrm{SDR}})$  & \textcolor{red}{\num{2.13e-2}}         & \textbf{\textcolor{greenbold}{\num{9.21e-1}}} &\num{9.28e-1}\\
$\mathrm{std}(\lambda_{\infty,\mathrm{SDR}})$& \textcolor{red}{\num{1.03e-2}}          &\textbf{\textcolor{greenbold}{\num{2.59e-1}}}  & \num{2.61e-1}\\
    \bottomrule
    \end{tabularx}
\end{table}
\begin{table}[t] \vspace{-7pt}
    \small
    \centering
    \caption{Use case scheduling dimension reduction guidelines} \vspace{-5pt}\label{tab: SDR_recommendations}
    \begin{tabularx}{1\columnwidth} { 
     >{\hsize=.6\hsize }X >{\hsize=.4\hsize }X}
     \toprule
    Use case & Method\\
    \midrule
    Large $n_x$ with moderate $n_p$ & DNN, Traj. PCA \\
    Large $n_x$ with large $n_p$ & Traj.~PCA \\
    Low $n_x$ with large $n_p$ & DNN \\
    Dominant linear dynamics & Traj. PCA, {SDR balred} \\
    \bottomrule
    \end{tabularx} \vspace{-5pt}
\end{table}
\vspace{-5pt}
\section{Conclusions \label{sec: conclusion}} \vspace{-8pt}
In this paper the performance of different SOR and SDR methods are analysed using three affine LPV-SS benchmarks. Concerning the SOR, it has been found that \emph{Moment Matching} is the only method able to show extrapolation capabilities on all the three benchmarks, and the only one capable of reducing unstable and/or non-minimal SS representations. However, other reduction methods are able to obtain a higher performance depending of specific model properties. Regarding the SDR, only \emph{SDR balred} is able to show extrapolation capabilities for the three benchmarks, but the performance of \emph{Trajectory PCA} or \emph{DNN} methods is better. As the analysis results show that the suitability of the reduction methods depends on the model to be reduced, some reduction guidelines are presented based on the model properties and reduction objectives. In addition, the presented guidelines can serve as a strategy for an iterative state order and scheduling dimension reduction for large order systems. For future research, the possibility of developing a joint state order and scheduling dimension reduction needs to be investigated.
\vspace{-15pt}
\section*{Acknowledgements \label{sec: acknowledgements}} \vspace{-8pt} The authors would like to thank Peter Seiler (University of Michigan) and Pascal Gahinet and Rajiv Singh (Mathworks) for fruitful discussions and technical advices. 
\vspace{-5pt}

\bibliographystyle{IEEEtran} 
\bibliography{LPV_reduction} 
\end{document}